\documentstyle[12pt]{article}
\def\ehat{\hat{e}}

\def\al{\alpha}
\def\be{\beta}
 
\def\ep{\epsilon}

\def\calA{{\cal A}} \def\calB{{\cal B}}  

\def\calD{{\cal D}}

  \def\calO{{\cal O}}

\def\calV{{\cal V}}

\newsavebox{\prooffig}

\def\shead#1{\parbigskipn {\bf #1} \parmedskipn} 

\def\del        {  \partial  }

\def\deldel#1   {  {\partial\over \partial #1}  }

\def\defint#1#2 {  \int_{#1}^{#2}  }

\def\half       {  {1\over 2}  }
\def\rootof#1   {  \left( #1 \right)^{1/2}  } 

\def\trace      {  \mbox{Tr}  }

\def\abs#1      {  \vert #1 \vert  }
\def\ie         {  {\it i.e.}      }
\def\evalat#1   {  \left\vert_{#1} \right. }

\def\where      { \mbox{where}\qquad }
\def\comma          {\, ,}
\def\period         {\, .}
\def\lsim    {\lower .65ex \hbox{\ $\stackrel{<}{\sim}$\ } }
\def\gsim    {\lower .65ex \hbox{\ $\stackrel{>}{\sim}$\ } }
\def\vecii#1#2      {  \left(\begin{array}{c}#1\\#2\end{array}\right)  }
\def\veciii#1#2#3   {  \left(\begin{array}{c}#1\\#2\\#3\end{array}
                     \right)  }
\def\veciv#1#2#3#4  {  \left(\begin{array}{c}#1\\#2\\#3\\#4
                                 \end{array}\right)  }
\def\vecfv#1#2#3#4#5 {  \left(\begin{array}{c}#1\\#2\\#3\\#4\\#5
                                 \end{array}\right)  }

\def\matrixii#1#2#3#4            {  \left(\begin{array}{cc}#1&#2\\#3&#4
                                       \end{array}\right) }
\def\matrixiii#1#2#3#4#5#6#7#8#9 {  \left(\begin{array}{ccc}#1&#2&#3\\
                                     #4&#5&#6\\#7&#8&#9\end{array}
                               \right)  }
\def\mativ#1#2#3#4               {  \left(\begin{array}{cccc}
                                       #1\\#2\\#3\\#4\end{array}\right) }
\def\matv#1#2#3#4#5              {  \left(\begin{array}{ccccc}
                                     #1\\#2\\#3\\#4\\#5\end{array}
                              \right)  }
\def\eqabegin         {  \begin{eqnarray}  }
\def\eqaend           {  \end{eqnarray}  }
\def\nn               {  \nonumber  }
\def\bracetwo#1#2     {  \left\{ \begin{array}{l} #1 \\ #2 \end{array}
                         \right.  }
\def\bracetwocases#1#2#3#4  {   \left\{ \begin{array}{ll} #1 &
                                 \qquad #2 \\
                                 #3 & \qquad #4 \end{array} \right.  }
\def\bracebegin#1     {  \left\{ \begin{array}{#1}   }
\def\braceend         {  \end{array}\right.   }
\def\parn              {  \par\noindent }

\def\parmedskip        {  \par\medskip  }

\def\parbigskipn        {  \par\bigskip\noindent  }
\def\parmedskipn        {  \par\medskip\noindent  }

\def\parag#1           {\paragraph{#1} \mbox{ }\parmedskip\noindent}
\def\boxit#1#2      {  \vbox{\hrule\hbox{ \hskip -4.1pt \vrule\kern3pt 

                     \vbox
                    {  \hsize #1 \strut\kern3pt #2 \kern3pt\strut  }
                       \kern3pt  \vrule} \hrule  } }
\def\centerbox#1#2  {  \mbox{  }\par\bigskip  \hfil \boxit{#1}{#2} \hfil
                       \par\bigskip\noindent }
\def\rightbox#1#2   {  \hfill\boxit{#1}{#2}  }
\def\leftbox#1#2    {  \boxit{#1}{#2}  }
\def\fullbox#1      {  \boxit{\textwidth}{#1}  }
\def\rightfigspacebegin  {  \par\noindent\begin{minipage}[t]{10cm}  }
\def\rightfigspaceend    {  \end{minipage}\par\noindent  }
\def\leftfigspacebegin   {  \par\noindent
                             \hspace*{10cm}\begin{minipage}[t]{6cm} }
\def\leftfigspaceend     {  \end{minipage}\par\noindent  }
\def\titleandfile#1#2   {  \begin{center}{\Large\bf #1}\end{center}
                            \par\begin{flushright} #2 \end{flushright}  }
\def\msection#1      {  \begin{center} \section{#1} \end{center}   }
\def\nsection#1      {  \let\boldface\bf \def\bf{} \section{#1}
                           \let\bf\boldface   }
\def\mnsection#1     {  \begin{center} \nsection{#1} \end{center}  }
\def\capsection#1    {  \let\boldface\bf \def\bf{\sc} \section{#1}
                           \let\bf\boldface   }
\def\mcapsection#1   {  \begin{center} \capsection{#1} \end{center} }

\newcommand{\nullify}[1]{}
\def\ie{{\it i.e. }}

\def\zbar{{\bar{z}}}

\def\bd{{\del\Sigma}}
\def\fdot{{\dot{f}}}

\def\ehat{{\hat{e}}}
\def\Ninv{N^{-1}}
\def\Dinv{D^{-1}}
\def\nuhat{\hat{\nu}}
\def\zetabar{{\bar{\zeta}}}
\def\zetatil{{\tilde{\zeta}}}

\def\taubar{{\bar{\tau}}}
\def\Tint{\int_0^Td\tau}

\def\ftil{{\tilde{f}}}
\def\ztil{{\tilde{z}}}
\def\abs#1{{|#1|}}
\def\bra#1{\langle #1 |}
\def\ket#1{| #1 \rangle}
\def\altil{{\tilde{\al}}}

\def\delbar {\bar{\partial}}
\def\zetaimunu{\zeta^i_{\mu\nu}}
\def\Xupmu{X^{\mu}}

\def\Xupnu{X^{\nu}}

\def\aprime{\alpha^{\prime}}

\def\zetaoneAB{\zeta^1_{AB}}
\def\zetatwoAB{\zeta^2_{AB}}

\def\ehatmu0{\hat{e}^{\mu}_0}
\def\ehatnu0{\hat{e}^{\nu}_0}

\def\nnu0theta{\nu_0 (\theta)}

\def\papertitlepage{\baselineskip 3.5ex \thispagestyle{empty}}
\def\Title#1{\baselineskip 1cm \vspace{1.5cm}\begin{center}
 {\Large\bf #1} \end{center} 
\vspace{0.5cm}}
\def\Authors#1{\begin{center} {\it #1} \end{center}}
\def\Abstract{\vspace{1.0cm}\begin{center} {\large\bf Abstract} 
           \end{center} \par\bigskip}
\def\Komabanumber#1#2#3{\hfill \begin{minipage}{4.2cm} UT-Komaba #1
              \parn #2 
              \parn #3 \end{minipage}}
\renewcommand{\thefootnote}{\fnsymbol{footnote}}
\renewenvironment{thebibliography}{\pagebreak[3]\par\vspace{0.6em}
\begin{flushleft}{\large \bf References}\end{flushleft}
\vspace{-1.0em}

\begin{enumerate}\if@twocolumn\baselineskip=0.6em\itemsep -0.2em
\else\itemsep -0.2em\fi\labelsep 0.1em}{\end{enumerate}}
\begin{document}
\papertitlepage
\vspace*{0cm}
\Komabanumber{96-26}{hep-th/9612064}{December 1996}
\Title{Scattering of Closed String  States \\ from a Quantized
 D-particle} 
\vspace{1cm}
\Authors{{\sc S.~Hirano\footnote[2]{hirano@hep1.c.u-tokyo.ac.jp; 
          JSPS Research Fellow.} 
 and Y.~Kazama
\footnote[3]{kazama@hep3.c.u-tokyo.ac.jp}
\\ }
\vskip 3ex
 Institute of Physics, University of Tokyo, \\
 Komaba, Meguro-ku, Tokyo 153 Japan \\
  }
\baselineskip .7cm
\Abstract
By developing an appropriate path-integral formalism, we compute, 
 in bosonic string theory, 
 the disk amplitude for the scattering of closed 
 string states from a D-particle, 
in which the collective coordinate of the D-particle 
 is fully quantized.  As a consequence, the recoil of the D-particle
 is naturally taken into account. Our result can be readily 
 factorized in the closed string channel to yield the 
  boundary state describing the recoiling D-particle. 
 This turned out to agree with the  BRST invariant 
 vertex  recently  proposed  by Ishibashi  to the leading order 
in the derivative expansion,
 but it will receive  corrections in 
 subsequent orders. The advantage of  our formalism is that it 
 is extendable to deal with more general processes 
 involving multiple  D-particles. A viewpoint regarding our work as 
 describing a dynamical transition of CFT's is also discussed.  
\newpage
\section{Introduction}
 D-branes have by now  acquired indisputable citizenship in the 
 world of string theory \cite{CJP}.  Born as  somewhat 
 exotic objects \cite{DGP}, their status was dramatically promoted 
 by the Polchinski's discovery \cite{Polch95}  that they are the 
 stringy representations of an important class of solitons 
 which carry Ramond-Ramond charges. Besides playing key roles 
 in numerous duality relations among string theories \cite{Polch96}
 and in the understanding of the statistical entropy of a black hole
 \cite{StromVafa}, they are  expected to provide  bridges  leading  
 us into the enigmanitc world of 11 dimensional M-theory
\cite{Townsend95,Witten953}\cite{BFSS}. 
 Thus their importance may well be even greater in  future developments. 
\par 
One of the many aspects of D-brane physics that need   
 better understanding is that of the dynamics of D-branes.  
 Indeed there have been a large number of investigations made on 
 this subject in the past year \cite{Bachas}\nocite{Lifschytz}\nocite{Barbon}
\nocite{KlebThor}\nocite{GHKM}\nocite{HK}\nocite{GM}\nocite{Witten9510}
\nocite{DanFerSun}\nocite{KP}-- \cite{MLI},
  which may be classified into 
 several categories according to the view points and methods
 employed. 
\par
To one category belong studies of the interaction of a D-brane 
  with another D-brane or with elementary string states 
 by means of open string
calculations \cite{Bachas}\nocite{Lifschytz}\nocite{Barbon}
\nocite{douglas}\nocite{KlebThor}
\nocite{GHKM}\nocite{HK}-- \cite{GM}.  
 They revealed 
 the nature of the velocity dependence of the inter-D-brane forces, 
 the internal structure of the D-brane through its 
 form factor and the decay characteristics, and many other 
 interesting properties. In these calculations, the D-branes 
 are treated as infinitely heavy background objects and hence the recoil 
 effects have so far not been taken into account. 
\par
Closely related are the investigations from the 
 closed string channel. 
By the duality property of the string worldsheet, open string 
 calculations  can be reinterpreted in the closed string 
 language, yielding a useful notion of boundary states \cite{PolchCai,CLNY}. 
 Using the T-duality transformations,  the components of the background 
 gauge field strength transverse to the D-brane worldvolume 
 can be interpreted as those of the velocity 
 of the D-brane and for constant velocity one can construct 
 the boundary state for a D-brane moving along a straight line. 
 This yields a complimentary view of the inter-D-brane
forces \cite{CalKleb,MLI}. 
Just as in the open string calculations, however, collective coordinates 
 of the D-brane are not quantized in such a treatment.  
\par
The third category is rather different in spirit 
 from the above two. 
It comprises the studies of low energy 
 interactions of D-branes using the dimensionally reduced 
 super Yang-Mills theory as the effective theory \cite{Witten9510}
\nocite{DanFerSun}\nocite{KP}-- \cite{DKPS}. 
The D-brane collective coordinates appear as the Higgs fields 
 on the worldvolume and are thus dynamical. 
This framework has been successfully employed for uncovering the 
bound states and resonances of D-branes. 
It can also be used for the scattering of D-branes  provided that 
 D-branes are nonrelativistic. Furthermore, it has been noted that 
 the 11 dimensional Planck scale naturally makes its appearance   
 in the regime treated in this formalism. 
\par
Although many intriguing characteristics of D-brane dynamics 
 have been uncovered through these studies, we are still at a 
 primitive stage, lacking in particular a more systematic formalism to
 deal with the collective dynamical degrees of freedom of the 
 D-branes themselves. In the first two categories sketched above, 
 D-branes are treated  as backgrounds and hence not fully dynamical.
 Attempts have been made to treat the recoil effects  
 within this type of setting,  with only a moderate success \cite{FPR,PT,KMW}. 
 The situation is  improved in the effective theory approach of the third 
 category, but its applicability is limited to the low energy 
 domain.  Moreover the connection between these two types of 
 approaches is understood only partially.    
\par
In this work, as a step toward more systematic treatment of D-brane
 dynamics, we  develop a path integral formalism in which the 
 collective coordinates of a D0-brane, \ie a D-particle,  are quantized. 
Specifically, we compute the disk amplitude for the  scattering of 
 closed string states from a quantized D-particle in bosonic 
 string theory.  
\par
The calculation is perfomed in two steps: First, in Sec.~2,  
 we set up a formalism with a D-particle moving along a fixed 
yet arbitrary  trajectory parametrized as $f^\mu(t)$, where $t$ is a worldline
coordinate. The effect of  the bulk fluctuations of the open string 
 attached to the D-particle is computed to all order in $\al'$ in the 
 usual fashion. On the other hand,  to incorporate  the  fluctuation of the 
ends of the string along the trajectory,  we need to  
 employ a generally covariant expansion in the number of derivatives of $f^\mu(t)$, regarded at this stage as  a background field.  
 As the only scale in the problem is $\al'$, this  
 takes the form of  an $\al'$ expansion, though distinct from  the 
 usual sense.  With this setting, we 
 compute  the scattering amplitude with closed string tachyon states
 to the leading order in the above expansion.  In the course of the 
 calculation, one finds that the conformal invariance requires the 
 trajectory to satisfy the current conservation condition $k_\mu 
\fdot^\mu  =0$, where $k_\mu$ is the total momentum of the 
closed string states and $\fdot^\mu = df^\mu/dt$.  Another point 
 to be mentioned  is   that   the integration over the fluctuation of
 the ends of the string  produces a divergence.
It is  of such a form that it can be absorbed 
 by the  renormalization of $f^\mu(t)$  when  there are no vertex 
 insertions.  With  vertex insertions, a part of the divergence
 remains in the amplitude but this will turn out to vanish upon 
 quantization of the trajectory.  Aside from this piece, 
 the resultant amplitude can be readily factorized in the closed 
 string channel and we obtain the boundary state describing 
 the D-particle moving along $f^\mu(t)$. This turned out to coincide  
 with the particle-particle-string vertex  recently proposed 
 by Ishibashi \cite{Ishibashi}, 
 who started with an ansatz which depends only on $\fdot^\mu$ 
  and imposed BRST invariance. Our 
 treatment makes it clear  that in general the boundary state will receive 
 corrections involving higher derivatives of $f^\mu(t)$ in subsequent 
 orders.   
 \par
In the second step, described in Sec.~3, we perform the integration
 over the trajectory $f^\mu(t)$ itself. The calculation  requires  
 regularization, upon which the divergent piece mentioned earlier 
 drops out.  An  important 
 outcome  is that, as it should be the case, the current conservation 
 condition $k_\mu \fdot^\mu =0$  automatically comes out  
  when the initial and the final D-particles are on shell. This was  
 anticipated in \cite{Ishibashi}, where, however, the quantization  was 
 not 
 explicitly performed. The final result is a  relativistic scattering 
 amplitude with the recoil of the D-particle fully taken into 
 account.   As an application of our formalism, we compute 
 the amplitudes with two tachyons and with two gravitons. 
 Besides the usual closed string poles in the $t$-channel, these  
 amplitudes exhibit  poles in a peculiar channel, which for small 
 momentum transfer is likely to be interpretable as $s$-channel 
  excitations of an open string with a heavy mass attached  at the ends. 
 Our result for the two graviton case agrees with the one 
  in \cite{KlebThor} in the limit of infinitely heavy  D-particle. 
\par
As our framework is rather general, the present work 
 can be extended in  many  directions. Some of these possibilities will 
 be discussed in Sec.~4. Finally, we will close this article by advocating 
 an intriguing viewpoint which regards our work as describing a 
 dynamical transition of CFT's.  
\renewcommand{\thefootnote}{\arabic{footnote}}
\section{Scattering Amplitude with Fixed D-particle Trajectory}
\shead{The setup}
Let us begin by computing the amplitude for closed string
 states  scattering from a classical D-particle with a specified 
 trajectory
 as a preliminary step to full quantization. \par
In this article we exclusively deal with the amplitude due to 
 the disk topology.  Under certain conditions to be discussed later, 
conformal invariance will be seen to hold\footnote{Precisely speaking,
we will be able to maintain BRST invariance. See also the discussion
at the end of Sec.~4.} and this allows us to take 
 the string worldsheet to be a unit disk denoted by $\Sigma$. 
Let  $\theta$ be the polar angle describing its boundary $\bd$.
 A D-particle 
 is characterized by the condition that the ends of the open 
 string terminate on the worldline of the D-particle, which we 
 parametrize by $f^\mu(t)$. As the ends may terminate anywhere on 
 the worldline, the precise Lorentz-covariant condition for a D-particle
 should  be expressed as \cite{Leigh}
\eqabegin
 X^\mu(\theta) &=& f^\mu(t(\theta)) \qquad \mbox{on $\bd$}\comma 
\eqaend
where $X^\mu$ denote the open string coordinates and 
 the function $t(\theta)$ is arbitrary. This means that 
in the path integral formulation we seek, we must integrate
 over $X^\mu(z)$ in the bulk and over $t(\theta)$ on 
 the boundary. Thus the relevant amplitude is given by
\eqabegin
\calV\left(f^\mu, \left\{k_i\right\}\right) &=& {1\over g_s}
\int \calD X^\mu(z) \calD t(\theta)
 \delta(X^\mu(\theta) - f^\mu(t(\theta)) 
 e^{-S\left[X\right]}\prod_i g_s V_i(k_i) \comma \label{fixedamp}
\eqaend
where $g_s$ is the string coupling constant, 
\eqabegin
 S\left[X\right] &=& {1\over 4\pi \al'} \int_\Sigma d^2z
 \del_\al X^\mu \del_\al X_\mu 
\eqaend
is the open string action\footnote{We use  Euclidean worldsheet 
 and the space-favored Minkowski metric $\eta_{\mu\nu}=
 \rm{diag}\ (-,+,+,\cdots, +)$ for the target space. Also we  write 
 $X^\mu(z)$ for $X^\mu(z,\zbar)$, etc.. } 
 and $V_i(k_i)$ are the vertex operators
 for closed string states carrying momenta $k_i$.  As will become clear  
 later, closed string states can be factorized so that we will be able 
 to deal with any states on equal footing.  With this in mind, we will 
 take  tachyon emission vertices $V_i(k_i) 
 = \int d^2z_i e^{ik_i \cdot X(z_i)}$ for illustration purpose.  

Consider first the $t(\theta)$ integral. It is convenient to split it 
 into the one over the $\theta$-independent mode, to be 
 denoted by $t$,  and the one over  the remaining non-constant mode. 
Further, in order to preserve the general coordinate invariance 
 along the trajectory, we use geodesic normal coordinate 
 expansion \cite{NCE} for  $f^\mu(t(\theta))$
 around $f^\mu(t)$. For the present case of one dimensional 
 submanifold embedded in a flat space, it is easy to find  
\eqabegin
 f^\mu(t(\theta)) &=& f^\mu(t) + \fdot^\mu(t)\zeta(\theta) 
 + \half K^\mu\zeta(\theta)^2 \nn\\
&&+{1\over 3!}
\left(-{\fdot^\mu \over h}K^\nu K_\nu 
 -{3\over 2}{\dot{h}\over h}
 K^\mu +P^{\mu\nu} \del_t^3f_\nu  \right) \zeta(\theta)^3
+\calO(\zeta^4) \period
\eqaend
Here, $\zeta(\theta)$ is the normal coordinate, a dot stands for 
 $t$-derivative,   and $h(t)$, $K^\mu(t)$
and  $P^{\mu\nu}(t)$ are, respectively, the one-dimensional 
 induced metric on the trajectory, the extrinsic curvature and 
 a projection operator normal to the trajectory. Their explicit 
 expressions are  
\eqabegin
h &\equiv & \fdot^\mu \fdot_\mu\comma \\
K^\mu &\equiv & \ddot{f}^\mu -\half{\dot{h}\over h}  \fdot^\mu = P^{\mu\nu}
\ddot{f}_\nu\comma  \\
P^{\mu\nu}&\equiv & \eta^{\mu\nu} -h^{\mu\nu}\comma \\ 
 h^{\mu\nu} &\equiv&{\fdot^\mu\fdot^\nu \over h}\comma
\eqaend
where we have also introduced the projection operator $h^{\mu\nu}$ along 
 the trajectory . We will take the functional measure 
 $\calD t(\theta)$ to mean $dt \calD \zeta(\theta)$ in the following. 
For lack of means  to perform the $\zeta(\theta)$-integration exactly, 
 we will treat the non-Gaussian higher order corrections pertubatively. 
Since this can be regarded as an expansion in the number of derivatives 
 in  $t$,  the basic
 piture  of our treatment is that to the zero-th order the boundary
 of the disk is attached to a point $t$ (to be 
 integrated) on the trajectory and the effects of the non-local 
 spread  is then taken into account by 
 the subsequent integration over $\zeta(\theta)$.
 It should be emphasized that although we must assume the smoothness of 
 the trajectory in order to be able to truncate the expansion,  
 the approximation nevertheless is fully covariant. 
\par
As for  $X^\mu(z)$ we likewise split it into the constant and the 
 non-constant modes:
\eqabegin
X^\mu(z) &=& x^\mu + \xi^\mu(z)\period
\eqaend
Then the $\delta$-function at the boundary 
 decomposes  into the product 
\eqabegin
&& \delta (X^\mu(\theta) -f^\mu(t)-\fdot^\mu(t) \zeta(\theta) - \cdots)
\nn\\
&& \quad =\delta(x^\mu -f^\mu(t)) \delta(\xi^\mu(\theta)
 -\fdot^\mu(t) \zeta(\theta) - \cdots) \label{constbd}
\period
\eqaend
%
\shead{Integration over $X^\mu(z)$}
With this setup, we now perform the integration over $X^\mu(z)$.
 $x^\mu$ integral 
 trivially replaces the zero mode $x^\mu$ in 
 the vertex operators by $f^\mu(t)$, giving 
\eqabegin
 e^{ik^\mu f_\mu(t)} \label{eikf} \comma \\
\where k^\mu &\equiv & \sum_i k_i^\mu \period
\eqaend
 To perform the one over 
$\xi^\mu(z)$, it is convenient 
 to set up a complete orthonormal moving frame
 $\left\{ \ehat^\mu_A\right\}\comma \ A=(0,a)$, $a=1,2,\ldots D-1$, 
 where $D=26$ is the dimension of the target spacetime:
\eqabegin
\ehat^\mu_0 &=& {\fdot^\mu \over \sqrt{-h}}\comma  \\
\ehat^\mu_A \ehat_{B\mu} &=&\eta_{AB}\comma \\
 \ehat_A^\mu \ehat_B^\nu \eta^{AB} &=& \ehat^\nu_A \ehat^{\mu A}
 = \eta^{\mu\nu}=
 -\ehat^\mu_0 \ehat^\nu_0 +  \sum_a\ehat_a^\mu \ehat_a^\nu\comma  \\
\ehat_0^\mu \ehat_0^\nu &=& -h^{\mu\nu}\comma  \\
\sum_a\ehat_a^\mu \ehat_a^\nu &=& P^{\mu\nu}
=\eta^{\mu\nu} -h^{\mu\nu}\period
\eqaend
Then we can decompose $\xi^\mu(z)$ in this frame as 
\eqabegin
\xi^\mu(z) &=& \sum_A \ehat_A^\mu \rho^A(z)\comma \\
\rho_A(z) &=& \ehat^\mu_A\xi_\mu(z) \period
\eqaend
$\rho_0$ ($\rho_a$) describes the fluctuation in the tangential
( transverse) direction. At the boundary, the constraint
 (\ref{constbd}) keeps the fluctuations truly along the 
 curved trajectory. In terms of $\rho_A$, this reads 
\eqabegin
\rho_0(\theta) &=& -\sqrt{-h}\zeta(\theta) +{1\over 3!} {K^2
 \over \sqrt{-h}}\zeta(\theta)^3 + \calO(\zeta^4) \comma \\
\rho_a(\theta) &=& \ehat^\mu_a\left( \half  \ddot{f}_\mu \zeta(\theta)^2
 + {1\over 3!} \left(-{3\over 2}{\dot{h} \over h}\ddot{f}_\mu
 + \del_t^3 f_\mu\right)\zeta(\theta)^3 \right)
 + \calO(\zeta^4) \period
\eqaend
The action for $\xi^\mu(z)$ now  becomes 
\eqabegin
{1\over 4\pi \al'}\int_\Sigma d^2z \del_\al \xi^\mu \del_\al \xi_\mu &=& 
 {1\over 4\pi \al'}\left(-\int_\Sigma d^2z \rho_A \del^2 \rho^A 
+ \int_\bd d\theta \rho_A 
 \del_n \rho^A\right)\period \label{rhobd}
\eqaend
In order to separate the bulk and the boundary interactions, 
 the surface integral should vanish, namely either $\del_n\rho_A
 =0$ or $\rho_A=0$ on the boundary. Since our expansion scheme 
 is such that to the leading order the boundary $\bd$ is mapped 
 to a point on a trajectory, we must adopt the boundary condition 
 so that $\rho_0(\theta)$ can fluctuate only along the trajectory 
 while $\rho_a(\theta) $ cannot fluctuate. This leads to the  
 condition  
\eqabegin
 \del_n \rho_0(\theta) &=& 0 \qquad \mbox{Neumann}\comma \\
 \rho_a(\theta) &=& 0 \qquad \mbox{Dirichlet}\comma 
\eqaend
which is precisely the familiar boundary condition for a D-particle. 

Now let us integrate over $\rho_A(z)$. To this end
 we write the $\delta$-function constraint in the form
\eqabegin
\delta(\xi^\mu(\theta) -\fdot^\mu \zeta(\theta) -\cdots) 
&=& \int \calD \nu^\mu(\theta) \exp\left( i \int d\theta 
 \nu_\mu(\theta) ( \ehat^{\mu 0}\rho_0(\theta) -\fdot^\mu \zeta(\theta) 
 -\cdots)\right)\comma  \nn
\eqaend
where we have used the boundary condition $\rho_a(\theta) =0$. 
Again it is convenient to introduce the projected quantities 
$\nu^A \equiv \nu_\mu \ehat^{\mu A}$. Then the above expression 
 reduces to 
\eqabegin
&&\int \prod_A \calD  \nu^A(\theta) \exp\left( i\int d\theta \nu^0 
\rho_0\right)
\nn\\
&& \quad \cdot \exp\left(
 -i\int d\theta  \left( \nu^0(\sqrt{-h}\zeta+\cdots)+ \nu^a 
(\half \ehat^\mu_a
 \ddot{f}_\mu \zeta^2 + \cdots)\right) \right)\period \label{nuint}
\eqaend
It is important to note that all the functions of $\theta$ do not 
possess  
 constant modes since such modes have already been separated. 
\par
Now it is straightforward to perform the integration over $\rho_A$. 
Together 
 with the tachyon vertex insertions, the relevant integral is 
\eqabegin
I &=& \int \calD \rho_A \exp\left({1\over 4\pi \al'} \int d^2z \rho_A 
\del^2\rho^A 
 + i\int d^2z J_A \rho^A \right)\comma  \nn\\
 J_A(z) &=& \sum_i k_{iA} \delta^2(z-z_i) 
 +\delta_{A0}\delta(|z|-1)\nu_0(\theta)\comma 
\eqaend
where we have introduced the projected momenta 
$k_{iA} = k_{i\mu}\ehat^\mu_A$. 
The result is 
\eqabegin
I &=& \exp\left( {\al'\over 2} \int d^2z d^2z' 
J_A(z)G_A(z,z')J_A(z')\eta_{AA} \right) \nn\\
&=&  \exp\left( {\al'\over 2}\sum_{i,j} k_{iA} k_{jA}G_A(z_i, z_j)
\eta_{AA}
 \right)\nn\\
&& \cdot \exp\left(-{\al'\over 2}\int d\theta d\theta' \nu_0(\theta)
 N(\theta, \theta')
\nu_0(\theta') \right)\nn\\
&& \cdot  \exp\left( -\al'\int d\theta  \sum_i k_{i0}\tilde{N}(z_i, 
\theta) \nu_0(\theta)\right)\period \label{rhoint}
\eqaend
Here $G_A(z,z')$ are the Neumann and the Dirichlet functions 
on the unit disk given by 
\eqabegin
 \del^2  G_A(z,z') &=& 2\pi\delta^2(z-z') \comma \\
G_0(z,z') &=& N(z,z')= \ln |z-z'| + \ln |1-{1\over z\bar{z'}}| \comma \\
G_a(z,z') &=& D(z,z')=\ln |z-z'| -\ln |1-z\bar{z'}| \period
\eqaend
 $\tilde{N}(z_i, \theta)$ is  the Neumann function 
 with one argument on the boundary 
  with the  zero mode part  omitted.\footnote{ This follows from the remark 
made earlier,  in particular from the lack of zero mode for 
$\nu_0(\theta)$. Explicitly,  $\tilde{N} (z, \theta)= 
 N(z, \theta)-\ln (1/|z|)$ and it vanishes for $z=0$}  
 $N(\theta,\theta')$, the Neumann function with both arguments 
 on the boundary,  possesses no zero mode and has 
 a regularized representation \cite{FradTseyt,AT} 
 which will be needed to deal with the divergence at $\theta=\theta'$:
\eqabegin
N(\theta,\theta')&=& -2\sum_{n=1}^\infty {e^{-\ep n}\over n}
 \cos n(\theta -\theta') \period \label{regneumann}
\eqaend
It is easy to check that its inverse $N^{-1}(\theta, \theta')$ in the 
space of non-zero  modes on a circle is give by 
\eqabegin
&&  N^{-1}(\theta, \theta') = -{1\over 4\pi^2} \del^2_\theta 
N(\theta, \theta')\comma  \\
&& \int  d\phi  N(\theta, \phi) N^{-1}(\phi, \theta') 
 =\tilde{\delta}(\theta-\theta') \equiv  \delta(\theta -\theta') 
-{1\over 2\pi}\period
\eqaend
\par
At this stage  we perform the integration over $\nu_0(\theta)$.  
Assembling terms containing $\nu_0$ from (\ref{nuint}) and 
 (\ref{rhoint}), the relevant integral is   
\eqabegin
&&\int \calD \nu_0 \exp\left( {-\al'\over 2}\int d\theta d\theta' 
 \nu_0(\theta)N(\theta,\theta')\nu_0(\theta') +i\int d\theta j_0(\theta) 
 \nu_0(\theta)\right)\nn\\
&& =  \exp\left( -{1\over 2}\int d\theta d\theta' 
 \tilde{\jmath}_0(\theta)N^{-1}(\theta, \theta')\tilde{\jmath}_0
(\theta')  \right)
\label{nuzeroint}
\eqaend
where  we have absorbed the factor of $\al'$ and defined 
\eqabegin
\tilde{\jmath}_0(\theta) &\equiv&{j_0(\theta) \over \sqrt{\al'}} = 
 i\sqrt{\al'}\, \tilde{k}(\theta)+ \sqrt{-{h\over \al'}}\, \zeta(\theta)
 -{1\over 3!} {K^2\over \sqrt{-h\al'}}\zeta^3(\theta) +\calO(\zeta^4)
\comma \\
\tilde{k}(\theta) &\equiv & \sum_ik_{i0}N(z_i, \theta)\period
\eqaend
\shead{Integration over $\zeta$}
We are now ready to perform the $\zeta$-integration,  
which takes the form of an  $\al'$-expansion. As was already emphasized  in the introduction, this is simply an organization of the 
 derivative expansion  and should not be confused with the usual 
 $\al'$ expansion in the context of  non-linear $\sigma$ model. 
To facilitate this expansion, we  rescale the 
 normal coordinate and introduce $\bar{\zeta}$ defined by 
\eqabegin
 \zeta &=& \sqrt{{\al'\over -h}}\, \bar{\zeta} \label{resc} \period
\eqaend
The change of the functional measure due to this transformation 
can easily be computed by expanding $\zeta(\theta)$ into Fourier 
 modes and employing the $\zeta$-function regularization. 
Remembering that the zero mode has been removed, we get 
\eqabegin
 \calD\zeta &=& \calD\left( \sqrt{{\al'\over -h}}\, 
  \zetabar\right) =\left( \prod_{n=1}
^\infty \sqrt{{\al'\over -h}} \right)^2 \calD \zetabar \nn\\
&=& \exp\left( 2\ln \sqrt{{\al'\over -h}}\,
 \sum_{n=1}^\infty 1\right)\calD \zetabar
 = \sqrt{{-h \over \al'}}\,  \calD \zetabar \period
\eqaend
With the  rescaling (\ref{resc}),   $\tilde{\jmath}_0(\theta)$ becomes 
\eqabegin
 \tilde{\jmath}_0(\theta) &=&\bar{\zeta} (\theta) + i\sqrt{\al'}\,  
\tilde{k}(\theta) 
 -{1\over 3!} \al' {K^2 \over h^2}\bar{\zeta}(\theta)^3 
 +\calO(\al'^{3/2}) \label{jtil} \period
\eqaend
The integrand for the $\zeta$-integration consists of 
 (\ref{nuzeroint}) plus the boundary contribution due to 
 the transverse fluctuation. Introducing the variables 
$\nuhat^\mu \equiv \nu^a\ehat^\mu_a$ and taking into account 
 the rescaling (\ref{resc}),  the expression to be added to the 
 exponent of (\ref{nuzeroint}) is 
\eqabegin
 {i\al'\over 2h}\nuhat^\mu\ddot{f}_\mu \zetabar^2 +\calO(\al'^{3/2})
\period
\eqaend
Substituteing (\ref{jtil}) into (\ref{nuzeroint}), adding the above 
 contribution and 
 keeping  terms up to $\calO(\al')$, the $\zetabar$ integration to be 
 performed becomes 
\eqabegin
 \int \calD\zetabar \exp\left(- \half \bar{\zeta}D \bar{\zeta}
 -ij_\zeta  \bar{\zeta} +{\al'\over3!} {K^2\over h^2} \bar{\zeta} 
\Ninv {\bar{\zeta}}^3 
+{\al'\over 2} \tilde{k}\Ninv \tilde{k} \right)\comma \label{zetabarint}
\eqaend
where we have used condensed notations such as 
$\zetabar D\zetabar = \int d\theta d\theta' 
 \zetabar(\theta) D(\theta, \theta') \zetabar(\theta')$, etc.. 
$D$ and $j_\zeta$ appearing in this expression are defined as 
\eqabegin
D &\equiv & \Ninv -i\al'{\nuhat^\mu\over h} \ddot{f}_\mu 
 = \Ninv \left( 1-i\al'N{\nuhat^\mu\over h} \ddot{f}_\mu\right)\comma \\
 j_\zeta &\equiv & \sqrt{\al'}\Ninv \tilde{k}\period
\eqaend
To perform this integral, we make a shift $\zetabar = 
\zetatil -i\Dinv j_\zeta $.  This produces a term which cancels the 
 last term in (\ref{zetabarint}). 
Again keeping terms up to $\calO(\al')$, the exponent becomes 
\eqabegin
 - \half \zetatil D\zetatil +{\al'\over3!} {K^2\over h^2}\zetatil \Ninv
 \zetatil ^3 \period
\eqaend
The second term, quartic in $\zetatil$, will be treated perturbatively. 
Then the integration produces two terms. One is the 
 determinant factor   proportional to 
\eqabegin
\trace\left(i\al'N{\nuhat^\mu \over h} \ddot{f}_\mu\right)
&=&{i\al'  \ddot{f}_\mu\over h} \int d\theta N(\theta,\theta) 
\nuhat^\mu(\theta) \period \nn
\eqaend
With the regularization discussed before, $N(\theta,\theta)$ is 
 a (divergent) constant. The remaining $\theta$ integral of  
  $\nuhat^\mu (\theta)$
 vanishes since it does not contain the zero mode. Therefore, to 
 this order, $\nuhat^\mu$ integral is inert. 
\par 
The other contribution is due 
 to the quartic term. 
Contractions of $\zetatil$'s produce a  divergent contribution 
\eqabegin
&& {\al'\over 2}{K^2 \over h^2} N(\theta',\theta') \int d\theta 
d\theta'  N^{-1}(\theta, \theta') N(\theta, \theta')  \period
 \eqaend
As for $\int d\theta d\theta'  N^{-1}(\theta, \theta') 
N(\theta, \theta')$, we can employ $\zeta$-function regularization 
and find that it is actually finite:
\eqabegin
\int d\theta 
d\theta'  N^{-1}(\theta, \theta') N(\theta, \theta') 
&=& \int d\theta {1\over \pi} \sum_{n=1}^\infty 1 =-1 \period
\eqaend
On the other hand $N(\theta',\theta')$ is truly divergent and 
 using the regularized form given in (\ref{regneumann}) we get
\eqabegin
N(\theta', \theta') &=& 2\ln \ep + \calO(\ep) \period 
\eqaend
This divergence signals the breakdown of conformal invariance for 
 a general trajectory. 

Assembling all together (and appending a $1/g_s$ factor 
 indicating the disk amplitude), we find that to $\calO(\al')$ 
the $\zeta$-integration  yields 
\eqabegin
&&{1\over g_s} \sqrt{{-h\over \al'}}\, \left( 1-\al'{K^2 \over h^2}
\ln \ep \right) 
\period \label{bareaction}
\eqaend
If we recall the definition $h=\fdot^\mu \fdot_\mu$, the first factor 
 is nothing but the action of a relativistic particle of 
 mass  $1/(g_s \sqrt{\al'})$, expected of a D-particle, 
  and the second factor is the  correction produced by the 
 boundary interaction. We now show that the divergence produced above 
is precisely of such a form that it can be absorbed by the 
 renormalization (shift) of the collective coordinate $f^\mu(t)$
 and yields the renormalized action.  Let $f^\mu_R$
 denote the renormalized trajectory function and set 
\eqabegin
 f^\mu &=& f^\mu_R +\delta f_R^\mu \comma  \\
\where \delta f^\mu_R &=& -\al'{K_R^\mu \over h_R}\ln \ep 
\label{renormf} \period
\eqaend
Then by a simple calculation one can check that  $\sqrt{-h}$ becomes 
\eqabegin
 \sqrt{-h} &=& \sqrt{-h_R} + \delta \sqrt{-h_R} \nn\\
\delta\sqrt{-h_R} &=& {-1\over \sqrt{-h_R}} \fdot^\mu_R
 \delta \fdot_{R,\mu} = \al'{K^2_R \over h_R^2}\sqrt{-h_R}\comma 
\eqaend
showing that (\ref{bareaction}) simply becomes $\sqrt{-h_R/\al'}$. 
This is in parallel with the renormalization of the Dirac-Born-Infeld
 action discussed in \cite{AT}. 
\par 
Note that the $\beta$
 function read off from (\ref{renormf}) is proportional to 
 $K^\mu_R$ and if we set this to zero we obtain a straight line 
 classical trajectory. As we wish to describe a scattering which 
necessarily requires non-straight path, we cannot  set $K^\mu_R$
 to zero. This does not mean, however, that the consistency of the
theory is impaired. It will be shown that the on-shell amplitude will 
be fully BRST invariant. Further clarifying discussion will be given at the 
end of Sec.~4 
%
\shead{The amplitude and the condition on $f^\mu(t)$}
Putting everything together, the amplitude for the scattering of 
 tachyons from a D-particle becomes 
\eqabegin
\calV_T(f^\mu_R, \left\{ k_i\right\} ) 
 &=& {1\over g_s} \int dt  \sqrt{{-h_R \over \al'}} e^{ i k \cdot f }
\nn\\
&& \!\!\!\!\!\quad \cdot \int \prod_i (g_s d^2z_i) 
 \exp \left( {\al' \over 2} \sum_{i,j}{}'k_i^\mu k_j^\nu
 \left(  h_{\mu\nu}
 N(z_i, z_j) +   P_{\mu\nu} D(z_i, z_j) \right) \right) \label{ntachamp}
\!\!\comma 
\eqaend
where $k^\mu = \sum_i k_i^\mu$ is the total momentum of the 
 tachyons, and 
 the prime on the summation means that, as usual, we omit
 the singular part of the Green's functions for $i=j$. 
On the 
 right hand side, except for the factor $\sqrt{-h_R}$ just discussed, 
 $f^\mu(t)$ and its derivative are still bare quantities and 
 hence contain divergence. \par
First consider the factor $f^\mu k_\mu$
 in $e^{ik\cdot f}$. The divergent piece is proportional to 
\eqabegin
 k_\mu K^\mu_R &=& k_\mu P^{\mu\nu}_R \ddot{f}_{R\nu}
 = {d\over dt}(k_\mu \fdot^\mu_R) 
 -(k_\mu \fdot^\mu_R) {\fdot_{R\nu} \ddot{f}^\nu_R \over h_R} \comma \nn
\eqaend
which   vanishes if  the {\it conservation of the D-particle current }
\eqabegin
 k_\mu \fdot^\mu_R &=& 0
\eqaend
holds. \par
The same condition is seen to arise from the requirement of the 
 $SU(1,1) \simeq SL(2,R)$ invariance of the integrand. 
The $SL(2,R)$ transformation 
 of the unit disk can be written as 
\eqabegin
z \longrightarrow \ztil &=& {\al z +\be \over \bar{\be} z +\bar{\al}}
\comma  \\
\abs{\al}^2 -\abs{\be}^2 &=& 1\period
\eqaend
The factors containing a particular coordinate, say $z_1$, produced 
 by this transformation are  
\eqabegin
 F_1 &=& \abs{\bar{\be}z_1+\bar{\al}}^{-4} \comma \\
F_2&=& \left(\abs{\bar{\be}z_1+\bar{\al}}
\abs{\bar{\be}\zbar_1^{-1}+\bar{\al}}\right)^{-\al' k_{10}\cdot k_{20}} 
\abs{\bar{\be}\zbar_1^{-1}+\bar{\al}}^{-\al' k_{10}^2} \comma \\
F_3 &=& \abs{\bar{\be}z_1+\bar{\al}}^{\al' \sum_a k_{1a}^2}\period
\eqaend
$F_1$ is from the integration measure, $F_2$ from the Neumann 
 function  and $F_3$ from the Dirichlet function. Putting them 
 together, we easily see that they cancel if and only if the 
 following two conditions are met:
\eqabegin
0&=& k\cdot \fdot \comma  \\
0 &=& \al' \left( \sum_a k_{ja}^2 +k^2_{j0}  \right)-4 
= \al'k_j^2-4 \qquad \mbox{for all $j$}\period
\eqaend
The first is the  current conservation, as promised, and the 
 second is the on-shell condition for the tachyons. 

At this stage, the current conservation is as yet a condition 
that can only be imposed by hand for consistency. 
When the dynamics of the D-particle itself is 
 taken into account, however, we will see that it arises naturally.  
 \par
Next consider the $\ln\ep$'s  residing in $h_{\mu\nu}$ and 
$P_{\mu\nu}$ in the last exponential factor of (\ref{ntachamp}). 
Even with the current conservation, they remain in the 
 form $c_1 (\del_t^2 f_R(t))^2 \ln \ep + c_2 \del_t^3 f_R(t)\ln \ep$, 
 where $c_1,c_2$ are functions of $k_i, z_i$ and $\fdot_R$. 
These unwanted terms, however, will be seen to vanish by quantum 
averaging over the trajectory, to be performed in the next section. 
\par
With the divergent terms put aside, the rest of the calculation is 
 completely standard. As an illustration, 
 consider the case of two tachyons. The $SL(2,R)$ invariance 
 allows us to fix  the position of one of the tachyon, say $z_1$, 
( and  if desired the angle of $z_2$) to $0$ and the amplitude becomes 
\eqabegin
\calV_T(f^\mu_R, \left\{k_i\right\}) &= & g_s
  \int dt  \sqrt{{-h_
R\over \al'}} e^{ i k \cdot f_R }
 \int _0^1 dy y^{(\al'/2) k_1\cdot k_2} (1-y)^{\al' 
(k_1\cdot \fdot_R)^2/\fdot_R^2 - 2} \comma \label{twotachamp}
\eqaend
where on the right hand side  all the $f^\mu$'s  are  renormalized.
Hereafter we shall omit the subscript R throughout.  
\shead{Boundary state representation}
Looking at the expression (\ref{ntachamp}), one immediately notices 
 that the exponential in the second line is nothing but 
 the usual open string amplitude except that for the directions 
 normal to the trajectory the Neumann function is replaced by 
 the Dirichlet function. This means that we can  factorize the 
closed string vertices and  isolate the D0-D0-string interaction 
 vertex in the form of a boundary state as seen from the closed string 
channel.  \par
Let us demonstrate this for the $N$ tachyon amplitude. 
In order to compare with (\ref{ntachamp}), where $|z_i| \le 1$, 
 we should use the \lq\lq bra" boundary state prepared at 
$z=1$ given by 
\eqabegin
\bra{B;f}&=& \bra{f} \exp\left( -\sum_{n\ge 1}{1\over n} \al^\mu_n
 D^\nu_\mu \altil_{n,\nu} \right)\comma \\
 D_\mu^\nu &=& h_\mu^\nu - P_\mu^\nu \comma 
\eqaend
where $\bra{f}$ represents the vacuum for the non-zero modes and the 
 position eigenstate for the zero mode.\footnote{Since the zero mode
of $X^{\mu}(z)$ is identified with $f^{\mu}(t)$, we should take the
position eigenstate for {\it all directions}.}  
It satisfies 
\eqabegin
\bra{B;f} \al_{-n}^\mu&=& \bra{B;f} 
(-D^\mu_\nu \altil_{n,\nu})\comma \\
\bra{B;f} \altil_{-n}^\mu&=& \bra{B;f}
 (-D^\mu_\nu \al_{n,\nu})\comma \\
\bra{B;f} x^\mu &=&\bra{B;f}  f^\mu \period
\eqaend
It is then straightforward to get 
\eqabegin
&& \bra{B;f}e^{ik_1\cdot X(z_1,\zbar_1)}e^{ik_2\cdot X(z_2,\zbar_2)}
\cdots e^{ik_N\cdot X(z_N,\zbar_N)}\ket{0} \nn\\
&& = e^{ik\cdot f}e^{ {\al'\over 2} 
\left( \sum_{i,j}' k_i^\mu P_{\mu\nu}k_j^\nu D(z_i, z_j) +
\sum_{i,j}' k_i^\mu h_{\mu\nu}k_j^\nu N(z_i, z_j) 
 + 2\sum_i k_i^\mu\ln|z_i| h_{\mu\nu}k^\nu \right) } \comma 
\eqaend
where $k^\mu$ is, as before,  the total momentum of the tachyons. 
Note that this agrees with (\ref{ntachamp}) except for the 
 last term in the second exponent. This term vanishes {\it if and 
 only if the current conservation condition $k_\mu \fdot^\mu=0$
 is satisfied.} Thus with this condition, the amplitude 
with the  insertion of arbitrary 
 closed string vertices $V_i(k_i)$ can be written as 
\eqabegin
\calV(f^\mu,\left\{ k_i\right\} ) &=&{1\over g_s} 
 \cdot \int dt  \sqrt{{-h \over \al'}}
\int \prod_i (g_s d^2 z_i)  \langle B;f| 
 V_1(k_1,z_1)  V_2(k_2,z_2)\cdots V_N(k_N,z_N)\ket{0}\comma  \nn\\
\label{bsrep}
\eqaend
where  appropriate symmetrization for $V_i(k_i)$ is understood. 
If one wishes, one can of course add the ghost part of the boundary 
 state to make it into BRST invariant form. This was in fact the 
 guiding principle used by Ishibashi \cite{Ishibashi}. Our formalism   
 has given a firm ground for his proposal and at  the same time  
 revealed that in general  the boundary state receives  
  corrections involving higher derivatives of $f^\mu(t)$, just 
 like in the case of Dirac-Born-Infeld action \cite{AT}. 
\section{Quantization of D-particle and Amplitude 
 with Recoil}
\shead{Preliminary}
Having obtained the amplitude describing the interaction 
 of closed string states with a D-paraticle along a fixed trajectory, 
 we now quantize the trajectory itself. This means that  we 
 integrate over $f^\mu(t)$ with the weight $\exp( -im_0 \int_0^1 dt 
 \sqrt{-h})$, where $m_0 \sim 1/(g_s \sqrt{\al'})$ is  the bare 
 D-particle  mass  inversely proportional to the string coupling  $g_s$.
 The action in the exponent  is nothing but the previous  amplitude 
without  the vertex  operator insertions, as it should be for the first 
quantized  formalism \cite{Combinatorics} .  It is actually more 
convenient toturn it into the 
 Polyakov form. How this comes about  is well-known \cite{Polyakov}
 but for the sake of completeness let us briefly recall  the procedure. 
\par 
 First, one introduces 
 a Lagrange multiplier, call it $\al(t)$, so that the induced metric 
 $h$ can be  treated as an  independent variable. 
 Next, define  the proper time $\tau(t) \equiv \int_0^t dt'\sqrt{-h(t')}$
 and the  total proper time $T=\tau(1)$. Then the integration over $h$ 
 reduces to the one over $T$, the modulus of the trajectory. 
As was argued in \cite{Polyakov},  
the fact that  the fluctuation of  $\al(t)$ is short-ranged allows one
 to replace  it by  its mean value $<\al>\sim 1/\ep$, where 
 $\ep$ is a short-time cutoff. This  divergence and the one 
 arising  in the $T$-integration measure are then absorbed by the  
 mass renormalization, producing  the renormalized 
 mass term of the form $-\half m^2T$. This means that effectively the 
 original $\tau$ is replaced by $m\tau$ and the new $\tau$ 
 should be treated as a variable of dimension $\mbox{mass}^{-2}$.  

 The net result of all this is that, instead of $-m_0 \int_0^1 dt 
 \sqrt{-h}$, we may use the effective action of the Polyakov type, 
 namely 
\eqabegin
 S&=& \int_0^T d\tau \half (\fdot^2 -m^2) 
\eqaend
and perform an overall $T$-integration $\int_0^\infty dT$. {\it We will 
 take this as our starting point.}  Hereafter a dot denotes the 
 derivative with respect to $\tau$.
\shead{Integration over $f^\mu(\tau)$}
With this preliminary, we now begin the calculation  
 of the amplitude with quantized  D-particle.  Rather than 
 dealing with  an  amplitude with some definite closed string 
 vertex insertions, it is more advantageous to compute 
 the  amputated  vertex 
\eqabegin
 \langle \calB;  p^\mu, p'^\mu, k^\mu | 
 &=& (p^2+m^2)(p'^2+m^2)\int d^Df d^D f' e^{i(p\cdot f-p'\cdot f')} 
\int_0^\infty dT \nn\\
&& \int _{f^\mu(0)=f^\mu\atop  
 f^\mu(T)=f'^\mu}\calD f^\mu(\tau)
 e^{ i\int_0^T d\tau \half (\fdot^2 -m^2)} \int_0^T d\taubar e^{ ik\cdot 
f(\taubar)} \langle \tilde{B}; \del_\taubar^n f(\taubar)| \comma 
\eqaend
from which one can compute any scattering amplitude just as 
in (\ref{bsrep}). 
It describes the transition of a D-particle with the initial 
 momentum $p^\mu$ into one with the final momentum $p'^\mu$. 
Here $f^\mu$ and $f'^\mu$ are, respectively, the initial 
 and the final position of the D-particle, 
$k^\mu=\sum_i k^\mu_i$ is the total momentum of the closed 
 string states. We have also separated the 
 zero-mode part  so that  $\langle \tilde{B}; \del_\taubar^n f(\taubar)|
\comma  (n=1,2,3),  $ 
represents the non-zero mode part of the  boundary state for a fixed 
trajectory, which, in our approximation,  depends primarily  on 
$\fdot(\taubar)$, except for the divergent pieces involving  
 $\del_\taubar^n f(\taubar), (n=2,3),$ mentioned in the previous 
 section. 
\par
First, in order to isolate the effect of $f^\mu(\tau)$ integration 
 on the boudary state, we replace $\del^n_\taubar f(\taubar)$ in 
 the boundary state by  new variables  $v^\mu_n$ 
 via the insertion 
\eqabegin
1 &=& \prod_n \int dv_n \int d\omega_n \exp\left( i\omega_n\cdot 
 (v_n -\del_\taubar^n f(\taubar))\right)\period  \label{one}
\eqaend
Then the $f^\mu(\tau)$ integration to be performed is 
\eqabegin
\int \calD f^\mu \exp\left( i \Tint \half \fdot^2 -\sum_n i\omega_n 
 \cdot \del_\taubar^n f(\taubar) +ik\cdot f(\taubar)\right)\period
\eqaend
To extract the dependence on the endpoint values $f^\mu$ and 
 $f'^\mu$, we expand around the classical solution satisfying 
 the endpoint condition:
\eqabegin
 f^\mu(\tau) &=& f^\mu_{cl}(\tau)+ \tilde{f}^\mu(\tau)\comma  \\
f^\mu_{cl}(\tau)&=& {y^\mu \over T} \tau + f^\mu  \comma 
\eqaend
where $y^\mu = {f'}^\mu -f^\mu $ and the fluctuation satisfies 
$\tilde{f}^\mu(0) = \tilde{f}^\mu(T)=0$. 
The contribution from this classical part is  
\eqabegin
 && \exp\left( {i\over T} \left(\half y^2 +k\cdot y\taubar 
 -y \omega\right)+ik\cdot f\right) \period \\
\eqaend
Combined with the Fourier transform factor $e^{i p\cdot f-ip'\cdot f'}
=e^{-iy\cdot p' + if\cdot (p-p')}$, 
 the  integration over $f^\mu$ gives the momentum conserving 
 delta function $\delta(p+k -p')$, while that over 
 $y^\mu$ yields 
\eqabegin
 T^{D/2} \exp\left( -{i\over 2T}(\omega + p'T -k\taubar)^2\right)\period
\label{zmint}
\eqaend
\par
The remaining $\ftil$ integral to be performed is
\eqabegin
 \int_{\ftil(0)=\ftil(T) =0} \calD \ftil \exp\left( 
 {i \over 2} \Tint \dot{\ftil}^2 -\sum_n i\omega_n \cdot
\del_\taubar^n {\ftil}(\taubar)
 + ik\cdot \ftil(\taubar)\right) \label{ftilint} \period
\eqaend
This is easily done by writing $\ftil(\taubar) 
 = \int_0^T d\tau \delta(\tau -\taubar) \ftil(\tau)$. Then 
 the exponent of (\ref{ftilint}) becomes 
\eqabegin
 E_f &=& -{i \over 2} \int _0^T d\tau \ftil \del_\tau^2 \ftil 
 + i\int_0^T d\tau \ftil(\tau) j(\tau) \comma \\
\where j(\tau) &=& k\delta(\tau-\taubar) -\sum_n (-1)^n
 \omega_n \del_\tau^n \delta(\tau-\taubar) \period
\eqaend
Upon integrating over $\ftil(\tau)$, we get 
\eqabegin 
&& T^{D/2} e^{E\left[j\right]} \comma \\
&& E\left[j\right] = {i \over 2}\int d\tau d\tau' j(\tau) 
 G(\tau,\tau')j(\tau')\period \label{expj}
\eqaend
$T^{D/2}$ is the determinant factor and it cancels the similar 
 one produced by the zero mode integration in (\ref{zmint}). 
The Green's function $G(\tau,\tau')$ is defined to satisfy  
$\del_\tau^2 G(\tau,\tau')=\delta(\tau-\tau')$ and vanishes 
 at the endpoints of the interval $\left[0,T\right]$. 
Its explicit form is  
\eqabegin
G(\tau,\tau') &=& {1\over T} 
 \left[ \theta(\tau -\tau')(\tau -T)\tau' + \theta(\tau'-\tau)
 (\tau'-T)\tau \right] \comma 
\eqaend
where $\theta(x)$ is the usual step function with $\theta(0)\equiv 
\half$. Upon expanding $E\left[j\right]$ in (\ref{expj}), 
one is required to evaluate the expressions of the form 
$\del_\tau^n \del_{\tau'}^m G(\tau,\tau')|_{\tau =\tau'}$, 
 which in general contain $\delta(x)$ and its derivatives and 
 need regularization. If one takes the representation 
 $\delta(x) =(2/T)(\half +\sum_{n\ge 1}\cos (2\pi nx/T) )$ and employ 
 the standard $\zeta$-function regularization, one finds that 
 $\delta(x)$ and all its derivatives vanish at $x=0$. In this way
 one easily finds that the only non-vanishing expressions 
 of the above type are 
\eqabegin 
G(\tau,\tau)|_{\tau'=\tau} &=& {\tau \over T}(\tau-T) \comma \\
\del_\tau G(\tau,\tau')|_{\tau'=\tau} &=& {\tau\over T} -\half\comma  \\
\del_\tau\del_{\tau'}G(\tau,\tau')|_{\tau'=\tau} &=& 
 {1\over T}\period
\eqaend
This means that effectively contributions of $\del_\taubar^n \ftil
(\taubar)$ for $n\ge 2$ wash out to zero upon quantization of 
 the trajectory\footnote{This situation is expected to change
 if we take the higher order corrections into account 
 so that the particle action itself is modified.}. 
With these formulae, the exponent $E\left[j\right]$ is easily 
 evaluated to be 
\eqabegin
E\left[j\right]&=& i{ \omega^2 \over 2T} -i{k\cdot \omega \over T}
\left( \taubar -\half T\right) + {i \over 2T} k^2 \taubar(\taubar -T)
\period
\eqaend
Adding this to the contribution from the zero-mode integration
 (\ref{zmint}), we find 
\eqabegin
&& -{i \over 2T}(\omega + p'T -k\taubar)^2 \nn\\
&&  + i{ \omega^2 \over 2T} -i{k\cdot \omega \over T}\left( \taubar
 -\half T\right) + {i \over 2T} k^2 \taubar(\taubar -T) \nn\\
&& = -i\omega \cdot \left(p+{k\over 2}\right) + 
 i \half (p'^2 -p^2) \taubar -{i \over 2} p'^2 T\comma 
\eqaend
where  we used the momentum conservation. 
Note that the parts quadratic in $\omega$ have canceled out. 
Putting  the factor $e^{i\omega\cdot v}$ (see (\ref{one})) 
 back in, we see that the $\omega$-integration produces 
 a delta function 
\eqabegin
  \delta( v-(p+{k \over 2}))\period
\eqaend
Thus,  remarkably $v$ is completely determined to be equal 
 to $p+{k\over 2}=\half (p+p')$, the mean of the 
 initial and the final D-particle momenta. This result was 
 anticipated in \cite{Ishibashi} but now we have a proof. 
 As we can write 
\eqabegin
k\cdot v &=& k\cdot \left( p+{k\over 2}\right) 
 = \half (p'^2+m^2) -\half (p^2+m^2) \comma 
\eqaend
the crucial consistency condition , $k\cdot v=0$, is automatically
satisfied  when  
 the initial and the final states of the D-particle are 
 on shell. 

The remaining integrals over $\taubar$ and $T$ are trivial. 
They give 
\eqabegin
&& \int _0^\infty dT\int_0^T d\taubar e^{{i \over 2}(p'^2 -p^2)\taubar
 -{i\over 2}(p'^2 +m^2)T}  \nn\\
&& \propto {1\over p^2+m^2}{1\over p'^2+m^2}
\eqaend
\ie the D-particle propagator legs, to be removed for the proper 
 scattering amplitude. 
\par
Thus our final result in the form of the vertex is 
\eqabegin
 \langle \calB;  p^\mu, p'^\mu, k^\mu | &=& 
\bra{0} \exp\left( \sum_{n\ge 1}{1\over n} 
 \al_{n}\cdot \altil_{n} -2\sum_{n\ge 1}{1\over n} \al_{\mu,n}
 \altil_{\nu, n} {(p+p')^\mu (p+p')^\nu \over (p+p')^2}  \right) \nn\\
 && \times \delta(p+k-p')  \period
\eqaend
When appended with the ghost contribution it agrees with the one
proposed in \cite{Ishibashi}, for which BRST invariance was enforced
by construction. 

Let us make a brief remark on the $m\rightarrow \infty$ limit. 
In this limit, in $p$ and $p'$ the energy components dominate 
 and hence the last factor in the exponent of (94) tends to 
 $n^\mu n^\nu / n^2$ where $n^\mu=(1,0,\ldots, 0)$. This gives 
 precisely the boundary state for a stationary D-particle. 
Alternatively, we can examine this limit in the path integral 
 itself. If we Fourier-transform the amplitude back to the position 
 representation, we have a factor $\exp(i(f'-f)^2/2T-im^2T/2)$. 
 Thus as $m\rightarrow \infty$ the $T$-integral is dominated by small
 $T$ and this in turn forces $f_\mu' \sim f_\mu$, 
  showing that the D-particle does not move. 
\shead{ Simple  Applications}
As simple applications of our formalism, let us compute 
 the amplitude with two tachyons and  the one with two gravitons.  
\par
The two-tachyon on-shell amplitude is immediately obtained from 
  (\ref{twotachamp}) and is proportional to 
\eqabegin
\calA(p,p',k)_T &=& \int_0^1 dy y^{(\al'/2)k_1\cdot k_2} 
 (1-y)^{\al' (k_1\cdot v)^2/v^2 - 2} \nn\\
&\propto & 
 {\Gamma\left({\al'\over 2}k_1\cdot k_2+1\right)
\Gamma\left(\al' {(k_1\cdot v)^2\over v^2}-1\right)
\over \Gamma\left( {\al'\over 2}k_1\cdot k_2+
\al' {(k_1\cdot v)^2\over v^2}-1\right)} \comma \\
\where v&=& p+{k\over 2} \comma \qquad k_1^2=k_2^2 = {4\over \al'}
\period 
\eqaend
The first $\Gamma$-function in the numerator gives the usual $t$-channel 
 closed string poles at 
\eqabegin
 t &\equiv & -k^2 = -{4\over \al'}(1-n)\comma \qquad n=0,1,\ldots\period
\eqaend
On the other hand, the second  $\Gamma$-function has poles 
 in a peculiar channel, namely, 
\eqabegin
-{(k_1\cdot v)^2 \over v^2} &=&{1\over \al'}n\comma \qquad 
 n=0,1,\ldots \period  \label{pecpole}
\eqaend
Rewriting this in terms of  $s\equiv -(k_1+p)^2$ and $t$,  
 we find 
\eqabegin
 s &=& m^2 + {2\over \sqrt{\al'}}\sqrt{n} \sqrt{m^2-{t\over 4}} 
 -{4\over \al'} -{t\over 2} \\
& {\simeq \atop m^2 >> t} & m^2 +{2\over \sqrt{\al'}}m\sqrt{n}
  -{4\over \al'} -{t\over 2} +\calO\left({t\over \sqrt{\al'}m}\right)\period
\eqaend
Because of the presence of $t$ on the right hand side, they do  
 not represent genuin poles in the $s$-channel. However, for $t << m^2$,  
 we see  the excitation spectrum for which the energy scale    
 is given by the geometrical mean of  $m$ and the string scale
 $1/\sqrt{\al'}$  and the spacing is of square root type.  Intuitively, 
 they should represent the excitations of an open string with 
 a heavy mass attached at the ends. Although more detailed 
 study is required, they appear to be new objects  which can 
 only be seen in this type of relativistic treatment.
\footnote{These peculiar excitations have also been noted by
N.~Ishibashi and M.~Li \cite{pc}.} \par
%
As the second example,  consider the amplitude with 
 graviton insertions.  
The graviton vertices are given by
\begin{equation}
        V_{G}(k_i) = \int d^2
   z_i\zetaimunu\del\Xupmu(z_i)\delbar\Xupnu(z_i)
   e^{ik_i\cdot X(z_i)} \quad\mbox{with}\quad k_i^2 = 0,
        \label{gravvtx}
\end{equation}
where $\zetaimunu$ is a polarization tensor for a graviton and satisfies 
the conditions $
        k_i^{\mu}\zetaimunu = k_i^{\nu}\zetaimunu = 0$
  and $ \sum_{\mu}\zeta^i_{\mu\mu} = 0$. 
The calculation is facilated by 
  introducing the source term  $i J_{z\mu}(z)\del\Xupmu(z)
        + iJ_{\zbar\mu}(z)\delbar\Xupmu(z)$ in the string action. 
Then  the relevant amplitude can be expressed  as   
\begin{eqnarray}
        {\cal V}_{G} & = & \left.\prod_i\int d^2 z_i\zetaimunu
        \frac{\delta}{\delta J_{z\mu}(z^i)}\frac{\delta}{\delta 
  J_{\zbar\nu}(z^i)}F[J]\right|_{J=0}{\cal V}_T,
        \label{VGtot}
\end{eqnarray}
where ${\cal V}_T$ is the amplitude with tachyons and $F[J]$ is the factor containing the sources:
\begin{eqnarray}
        F[J]&=& \exp\!\left\{\!\aprime\sum_{i,j}\eta_{AA}
        \left( {1 \over 2}J_{\alpha A}(z_i)
        \del^{\alpha}\del^{\prime\beta}G_A(z_i,z_j)J_{\beta A}(z_j)
    -k_{iA}J_{\alpha A}(z_j)\del^{\prime\alpha}G_A(z_i,z_j)
        \right)\!\right\}. \nn\\
        \label{Jterm}
\end{eqnarray}
Here $\del^{\prime}$ denotes the derivative with respect to the second
 argument of the Green function.
 Restricting to the two graviton case 
 and performing the rest of the calculation in the gauge
$\zeta^i_{0A}=0$, we find the amputated  
 amplitude to  be proportional to  
\begin{eqnarray}
         & \displaystyle{{\alpha^{\prime 2} \over 8}
           \int_0^1dy\left\{\zetaoneAB\zetatwoAB
         \left( 1 + {1\over y^2}\right)
         +\aprime(\zetaoneAB k_{2B})(\zeta^2_{AC}k_{1C})
        \left( 1 - {1 \over y^2} \right)\right.} & \nn \\ 
         & \displaystyle{\left. + ({\aprime \over 2})^2
           (k_{2A}\zetaoneAB k_{2B})  (k_{1C}
         \zeta^2_{CD}k_{1D})\left( 1 -{2 \over y} + {1 \over
        y^2}\right)\right\}
         (1 - y)^{-\aprime k_{10}^2} y^{{\aprime \over 2}k_1\cdot k_2}} &. 
        \label{twograv}
\end{eqnarray}
Upon  $y$-integration, this becomes
\begin{eqnarray}
&&       ({\aprime \over 2})^3{\Gamma({\aprime \over 2}k_1\cdot k_2 -1)
         \Gamma(-\aprime k_{10}^2 + 1) 
         \over \Gamma({\aprime \over 2}k_1\cdot k_2 - \aprime k_{10}^2 + 2)}
          \left\{ \zetaoneAB\zetatwoAB \left( 
{\aprime \over 2}k_{10}^4 
   - k_{10}^2
         + {\aprime \over 2}\left(\sum_a k_{1a}k_{2a}\right)^2
\right)\right.     
  \nn \\ 
    &  & \qquad \qquad 
   -\aprime(\zetaoneAB k_{2B})(\zeta^2_{AC}k_{1C})
    \sum_a k_{1a}k_{2a}( 1 - \aprime k_{10}^2 ) \nn \\
  &  & \left.\qquad\qquad   + {\aprime \over 2}
   (k_{2A}\zetaoneAB k_{2B})(k_{1C}\zeta^2_{CD}k_{1D})
   (1 - {\aprime \over 2}k_{10}^2 )( 1 - \aprime k_{10}^2)\right\}.  
        \label{gravVeneziano}
\end{eqnarray}
 In the limit of infinitely heavy D-particle, this reduces to the 
 result obtained in \cite{KlebThor}. Just as in the previous example, 
   poles occur in  the  exotic channel. 
\section{ Discussions}
In this work,  we have initiated a path integral formalism for 
 the quantization of  D-particles.  Although  this article   
is devoted to explaining the basic formalism and 
 its applications in the simplest setting, our work should serve 
 as a starting point for investigations of a variety of 
 important problems, both conceptual and technical. Below 
 we shall discuss some of these perspectives.
\par
Let us begin with some immediate extensions.  
One obvious and necessary task is to extend the formalism to the 
 superstring case. This extension  will be reported elsewhere. 
\par
Another urgent application is the calculation of the 
 scattering amplitude for two quantum D-particles. This is important 
 in  many respects. In particular, this would allow us to 
 compare our apporach  with that using the low energy effective
  gauge theory 
  and clarify both the  foundation and the  limitation of the latter. 
 The key would be to understand the mechanism of the appearance of 
 the enhanced gauge symmetry and its spontaneous breakdown. 
 These matters are currently under investigation and we hope to report 
 our progress in the near future. 
\par
 Once the two-particle case is understood, the 
 next task will be to extend it to the case involving
  multiple D-particles. 
 Here, we expect to be able to make contact with the 
 extremely interesting proposal recently made by Banks et al \cite{BFSS}, 
 namely the formulation of the M-theory in terms of D-particles
 in the infinite momentum frame. One of the crucial questions 
 concerning this proposal is whether one can find the 
 action which is covariant with respect to the 11 dimensional 
 Lorentz group. As our path integral formalism (when 
 extended to the superstring) respects covariance 
 at least in the 10 dimensional sense, it may provide a clue 
 to this important question. 
\par
Aside from applications to D-particle systems, our formalism 
 can in principle be adapted to the processes involving D-strings,
 provided they are compactified to carry finite masses. 
 This might shed some light on the nature of 12 dimensional 
 F-theory \cite{Vafa}. 
\par
Finally, we wish to discuss a possible conceptual implication that 
 our work may have upon the outstanding problem of the 
 vacuum selection in string theory. From the standpoint of non-linear 
 sigma model, one consistent background corresponds to one 
 conformal field theory (CFT) and the remarkable discovery of 
 Polchinski \cite{Polch95} is that D-branes provide exact CFT's of
 that kind.  
 To make our discussions concrete, let us  focus upon the process  
  considered in this work.  Form a wavepacket for the  
 D-particle which is  initially moving along a straightline. As this 
 is a solution of the condition for the vanishing $\beta$-function, it 
 gives a CFT.  Similarly, the final packet moving along 
 a different straightline gives another CFT.  Therefore the scattering 
 process describes {\it a transition between two different CFT's.}  Moreover, 
 the quantum fluctuation necessarily involves non-CFT stages  
 in the middle.  Thus, strictly speaking, the conformal invariance 
 is not respected in the usual sense. Indeed,  to describe the scattering  
 we could not set the $\beta$-function to zero. This, however, does 
 not imply a disaster: Consistency of the theory is maintained  in 
 the form of BRST invariance.  This is somewhat reminiscent of the 
 situation that occurs in  non-linear sigma model, 
 where string loop corrections  effectively modify the $\beta$-function,
  which in turn can be derived from the requirement of BRST 
 invariance \cite{Lovelace,FS,CLNY2}. 
 As in many situations in string theory, the imperative requirement 
 is the unitarity and BRST invariance is the most powerful way to 
 implement this crucial consistency condition. 
\par 
Thus, to sum up, our work may be interpreted as an instructive example 
in which the dynamical transition between CFT's is consistently 
 described: Configuration space of D-particles can be regarded as 
 the  moduli space of a class of CFT's and,  upon  quantization,  
 points of this moduli space are dynamically connected.  It should be 
 extremely interesting to explore the implication of this view point 
 for the problem of vacuum selection in string theory and beyond.
\par\bigskip\noindent
{\large\bf Acknowledgment}\par\smallskip\noindent
We would like to acknowledge stimulating discussions with 
 M.~Kato and T.~Yoneya. We are also grateful to the 
 organizers and the participants of Kashikojima workshop for 
 providing us with a thought-provoking atmosphere. In particular, 
 we would like to thank N.~Ishibashi for explaining 
 his work \cite{Ishibashi} during the workshop. The work of S.~H. was
 supported in part by the Japan Society for the Promotion of Science. 

\newpage

\end{document}